\documentclass[preprint2, longabstract]{aastex6}

\usepackage{natbib}
\usepackage{tabularx}
\usepackage{array}
\usepackage{url}
\usepackage{hyperref}

\shorttitle{SETIBURST}
\shortauthors{Chennamangalam et al.}

\begin{document}

\title{SETIBURST:\\A Robotic, Commensal, Realtime Multi-Science Backend\\for the
    Arecibo Telescope}

\author{Jayanth Chennamangalam}
\affil{Astrophysics, University of Oxford, Denys Wilkinson Building, Keble
    Road, Oxford OX1 3RH, UK}
\email{jayanth@astro.ox.ac.uk}

\author{David MacMahon}
\affil{Department of Astronomy, University of California Berkeley, Berkeley,
    CA 94720, USA}

\author{Jeff Cobb}
\affil{Department of Astronomy, University of California Berkeley, Berkeley,
    CA 94720, USA}

\author{Aris Karastergiou\altaffilmark{1,2}}
\affil{Astrophysics, University of Oxford, Denys Wilkinson Building, Keble
    Road, Oxford OX1 3RH, UK}

\author{Andrew P. V. Siemion\altaffilmark{3,4}}
\affil{Department of Astronomy, University of California Berkeley, Berkeley,
    CA 94720, USA}

\author{Kaustubh Rajwade}
\affil{Department of Physics and Astronomy, West Virginia University,
    PO Box 6315, Morgantown, WV 26506, USA}

\author{Wes Armour}
\affil{Oxford e-Research Centre, University of Oxford, Keble Road,
    Oxford OX1 3QG, UK}

\author{Vishal Gajjar}
\affil{Department of Astronomy, University of California Berkeley, Berkeley,
    CA 94720, USA}

\author{Duncan R. Lorimer\altaffilmark{5}}
\affil{Department of Physics and Astronomy, West Virginia University,
    PO Box 6315, Morgantown, WV 26506, USA}

\author{Maura A. McLaughlin}
\affil{Department of Physics and Astronomy, West Virginia University,
    PO Box 6315, Morgantown, WV 26506, USA}

\author{Dan Werthimer}
\affil{Department of Astronomy, University of California Berkeley, Berkeley,
    CA 94720, USA}

\and

\author{Christopher Williams}
\affil{Astrophysics, University of Oxford, Denys Wilkinson Building, Keble
    Road, Oxford OX1 3RH, UK}

\altaffiltext{1}{Department of Physics and Electronics, Rhodes University,
    PO Box 94, Grahamstown 6140, South Africa}
\altaffiltext{2}{Physics Department, University of the Western Cape,
    Cape Town 7535, South Africa}
\altaffiltext{3}{ASTRON, PO Box 2, 7990 AA Dwingeloo, The Netherlands}
\altaffiltext{4}{Department of Astrophysics, Radboud University, PO Box 9010,
    6500 GL Nijmegen, The Netherlands}
\altaffiltext{5}{National Radio Astronomy Observatory, PO Box 2, Green Bank,
    WV 24944, USA}

\begin{abstract}
Radio astronomy has traditionally depended on observatories allocating time to
observers for exclusive use of their telescopes. The disadvantage of this
scheme is that the data thus collected is rarely used for other astronomy
applications, and in many cases, is unsuitable. For example, properly calibrated
pulsar search data can, with some reduction, be used for spectral line surveys.
A backend that supports plugging in multiple applications to a telescope to
perform commensal data analysis will vastly increase the science throughput of
the facility. In this paper, we present `SETIBURST', a robotic, commensal,
realtime multi-science backend for the 305-m Arecibo Telescope. The system
uses the 1.4~GHz, seven-beam Arecibo $L$-band Feed Array (ALFA) receiver
whenever it is operated. SETIBURST currently supports two applications:
SERENDIP VI, a SETI spectrometer that is conducting a search for signs of
technological life, and ALFABURST, a fast transient search system that is
conducting a survey of fast radio bursts (FRBs). Based on the FRB event rate and the
expected usage of ALFA, we expect 0--5 FRB detections over the coming year.
SETIBURST also provides the option of plugging in more applications. We outline
the motivation for our instrumentation scheme and the scientific motivation of
the two surveys, along with their descriptions and related discussions.
\end{abstract}


\section{Introduction}

Radio astronomy relies on observations for which telescope time was obtained
following a competitive proposal review. This process is critical because
telescope time is limited: Only one kind of observation can usually be done at
a given time. In addition to this exclusivity, the utility of the collected
data is usually restricted to the specific kind of experiment that it was
obtained for. Data reuse within a given field is standard practice -- e.g., the
original fast radio burst (FRB; see \S\ref{intro_frb}) was discovered in a
reprocessing of data from a fast radio transient survey of the Magellanic
Clouds using the 64-m Parkes Radio Telescope \citep{lor2007} -- but
cross-disciplinary data reuse is a rarity. For example, spectral line surveys,
due to long integration times used in its observations, result in data products
that cannot be reused in a search for pulsars. On the other hand, properly
calibrated pulsar search data can be used for spectral line surveys as well,
but it is rarely done. This severely restricts the science throughput of a
facility. To optimize data collection and analysis, commensal observing is
increasingly being employed, wherein multiple data processing/recording
processes run simultaneously on data from the telescope during observations. In
such a scheme, telescope pointing remains under the control of the primary
observer, but secondary observers also have access to data, vastly increasing
the available sky coverage.

Commensal observing was pioneered by the early searches for extraterrestrial
intelligence (SETI) at the Hat Creek Radio Observatory \citep{bow1983}, and
later, at the Arecibo Observatory, home to the 305-m diameter Arecibo
telescope. The original need for commensal observing was due to the fact that
SETI requires searching a large parameter space for which a significant amount of
telescope time is required, and the inability of allocating dedicated time to
such a large survey that is speculative in nature. The Search for Radio
Emissions from Nearby Developed Intelligent Populations (SERENDIP) project at
the Arecibo Observatory -- of which the instrument described in this paper is a
part -- has, throughout its existence, relied on commensal data processing
\citep[see, for example,][]{bow1993}. Technologically, in recent times,
relatively inexpensive networking hardware and high-performance computing (HPC)
machines have made it possible to build multiple HPC-based backends that can
easily distribute and process radio telescope data, enabling commensal data
processing. The Allen Telescope Array was built with commensal observing as a
design goal, such that SETI and non-SETI observations could be done in parallel
\citep{wel2009}. In high time resolution astronomy, the need for commensal
observing has been made apparent by the discovery of new classes of fast radio
transient, such as rotating radio transients \citep[RRATs;][]{mcl2006} and FRBs
\citep{lor2007,tho2013}. The V-FASTR experiment at the Very Large Baseline
Array \citep{way2012} is a commensal search for fast transients.
VLITE\footnote{\url{http://vlite.nrao.edu/}} is a ten-antenna system at the
Very Large Array that performs ionospheric observations, transient searches,
and imaging in parallel with regular observations. Among new facilties, the
Australian Square Kilometre Array Pathfinder (ASKAP) is used for the Commensal
Realtime ASKAP Fast Transients survey \citep{mac2010}. In this paper, we
describe a new instrument at the Arecibo Observatory that is centered around
the idea of commensal observing, with a SETI experiment and a fast transient
survey as consumers of the collected data.

The outline of this paper is as follows: In the following subsections, we
introduce the two science motivations of the project, namely, SETI and fast
radio transients. In \S\ref{sec_sysdesc}, we describe
the technical details of the system: the SERENDIP VI SETI backend and the
ALFABURST fast transient backend. In \S\ref{sec_surv}, we describe the two commensal
surveys we are undertaking, and conclude in \S\ref{sec_conc}.

\subsection{The Search for Extraterrestrial Intelligence}

The quest for life in the Universe has seen much progress in recent years, with
the exploration of the Solar System and the detection of a large number of
extrasolar planets. A whole new field -- astrobiology -- has emerged to tackle
the problem of whether life exists elsewhere in the Galaxy. SETI aims one step
further, to answer the more challenging question of the existence of
technological intelligent life. One of the first SETI attempts followed the
suggestion by \citet{coc1959} that ETI may transmit narrow-band beacons near
the radio emission line of neutral hydrogen, at 1420~MHz, enabling radio
astronomers in other civilizations to detect them. Radio SETI observations
started with \citet{dra1961} who searched for narrow-band lines, and have
continued to the present day with increasing levels of sophistication. For
instance, \citet{sie2013} recently searched for interplanetary radar signals in
multi-planet systems during conjunctions\footnote{Note
the error in the sensitivity calculation in \citet{sie2013}, and the values
reported in their Table 3: Their characteristic sensitivity should be
$\sim$2$\times10^{-22}$~erg~s$^{-1}$~cm$^{-2}$, and the exponent in footnote (d)
in their Table 3 should be $-$22.}, and established that
$\lesssim$1\% of transiting exoplanet systems host civilizations that emit
narrow-band radiation in the 1--2~GHz band with an equivalent isotropically
radiated power (EIRP) of $\sim$1.5$\times10^{13}$~W.

Whether ETI would set up beacons for the benefit of curious radio astronomers
in other civilizations is unknown, but setting up such beacons is the best
possible way to advertise our presence in the Universe. Radio emission is
energetically, and hence, economically inexpensive to generate. Radio waves can
travel vast distances with relatively less attenuation due to the interstellar
medium (ISM) compared to electromagnetic radiation at other wavelengths,
ensuring a better likelihood of signal reception. Cleverly-designed beacons,
such as extremely narrow band signals near a natural emission line commonly
used in studying the Galaxy, such as that proposed by \citet{coc1959}, will
increase the likelihood of the signal being noticed by radio astronomers
elsewhere. The rationale behind searching for extremely narrow-band signals is
that the narrowest astrophysical lines are of the order of hundreds of Hz wide.
The narrowest line detected has a width of 550~Hz \citep{coh1987}. Even if a
civilization does not set up a beacon, radio emission created by their
technology could leak out into space, at least during the early stages of their
technological development, in much the same way as coherent radio emission
produced quite commonly by human technology routinely leaks out to space. Given
that humans have been transmitting in the radio for more than a century,
the earliest radio transmissions have traversed a distance $>30$~pc away from
us. It is conceivable that leakage signals from other civilizations may be
picked up on Earth, although given the fact that they are not intentional
transmissions to us, it is unlikely that they would be easy to detect amid the
noise background.

\subsection{Fast Radio Bursts\label{intro_frb}}

Fast radio transients have been a staple of radio astronomy research for
decades, starting with the discovery of the first pulsars \citep{hew1968}. In
recent years, new classes of such transient have emerged, namely, RRATs, which
are thought to be highly intermittent pulsars \citep{mcl2006}, and FRBs
\citep{lor2007,tho2013}. However, in spite of the long history and recent
discoveries, robotic surveys in the radio are only beginning to be employed.
For example, the Survey for Pulsars and Extragalactic Radio
Bursts\footnote{\url{https://sites.google.com/site/publicsuperb/}} (SUPERB)
at the Parkes telescope uses the `Heimdall' realtime data processing pipeline.
A robotic system in the radio, that operates as long as a supported receiver is
available, can radically increase the time available on the sky and lead to
discoveries of fast transients such as RRATs and FRBs. Realtime detection has
the advantage of being able to trigger other telescopes that are geographically
separated and operate at other frequencies, as exemplified by \cite{kea2016},
whereas offline analysis of data usually results in a latency of days to
months. The ability to follow up detected events within hours or days is
critical to
the identification of potential afterglows or other indicators of the same
event at different wavelengths, helping shed light on the location of, and the
physics behind these exotic sources.

FRBs are broadband radio pulses with observed widths of the order
of a few milliseconds. Due to the nature of the dispersion caused by the
ionized ISM, lower frequency components of the pulse are delayed
much more than the higher frequency components. This delay is quantified in
terms of the dispersion measure (DM), which, for FRBs, is greater than that
contributed by the ISM of the Galaxy, indicating an extragalactic origin. The
millisecond time scale of the pulse implies a compact object progenitor.
Due to the fact that all reported FRBs have been detected using
single-dish radio telescopes that have wide beams, localization, and hence,
association with sources at other wavelengths, has proven to be a challenge.
Therefore, a conclusive explanation of what FRBs are has remained elusive,
with various theories being proposed. Extragalactic-origin theories that posit
cataclysmic explosions include the gravitational collapse of supramassive
neutron stars \citep{fal2014} and binary neutron star mergers \citep{tot2013}.
Extragalactic-origin theories that predict repetition include
giant-pulse-emitting pulsars \citep[see, for example,][]{lyu2016}, flaring
magnetars \citep[][for instance]{pop2013}, and Alfv\'en wings around planetary
companions to pulsars \citep{mot2014}. There is at least one Galactic-origin
theory (that also predicts pulse repetition) wherein these bursts originate in
flare stars, and are dispersed by the stellar corona \citep{loe2014}. A
resolution of the mystery has been made complicated by two recent discoveries,
namely, that of FRB 150418 which has been claimed to be associated with a slow
radio transient in an elliptical galaxy interpreted to be an afterglow of a
cataclysmic, non-repeating event \citep{kea2016}, and that of FRB 121102 which
has been shown to repeat \citep{spi2014,spi2016}\footnote{In
developments published during the late stages of the review process of this
manuscript, the repeating source FRB 121102 has been shown to be associated
with an extragalactic source \citep{cha2017}.}.

Eighteen
FRBs have been reported in published
literature\footnote{\url{http://www.astronomy.swin.edu.au/pulsar/frbcat/}}
\citep{pet2016}. All reported FRBs except two were discovered using the Parkes
telescope. The exceptions were discovered using the Arecibo telescope
\citep{spi2014} and the Green Bank Telescope \citep{mas2015}. Since FRBs are
non-repeating/highly intermittent, increasing the amount of observing time
available will allow more to be detected, enabling a better understanding of
these events.

\subsection{SETIBURST}

Keeping in mind the aforementioned scientific motivations, we have designed,
developed, and deployed SETIBURST, an automated, commensal, realtime
multi-science backend for the Arecibo telecope. SETIBURST has two plug-in
applications: SERENDIP VI
(referred to as `S6' henceforth), the latest in the SERENDIP series of SETI
spectrometers, and ALFABURST, a fast transient
search pipeline. SETIBURST performs commensal processing of signals
from the 1.4~GHz seven-beam Arecibo $L$-band Feed
Array\footnote{\url{http://www.naic.edu/alfa/}} (ALFA) receiver.
ALFA is Arecibo's workhorse survey receiver, being used for such large-scale
surveys as the Pulsar ALFA (PALFA) survey \citep{cor2006}, which has so far
resulted in the discovery of 145 pulsars and one FRB
\citep{cor2006,laz2015,spi2014};
the Galactic ALFA Continuum Transit Survey
\citep[GALFACTS; see, for example,][]{tay2010}; and the Arecibo Galaxy
Environment Survey \cite[AGES; see, for example,][]{aul2006}.
ALFA is well-suited for an FRB survey at Arecibo given that all except the
Green Bank FRB were detected at 1.4~GHz. Being a survey receiver with multiple
beams, ALFA is suited for SETI surveys as well.

\section{System Description}\label{sec_sysdesc}

SETIBURST is a heterogeneous instrument, i.e., it uses Field Programmable Gate
Arrays (FPGAs) and HPC machines equipped with
Graphics Processing Units (GPUs). Heterogeneous instruments have increased in
popularity in astronomical signal processing applications in recent years
\citep[see, for example,][]{dup2008,woo2010}, as they combine the high bandwidth
capabilities of FPGAs with the features, flexibility, and ease of programming
of GPUs. The network switch has the potential to act as a data hub into which
HPC backends may be plugged in, enabling multiple simultaneous experiments.

\begin{figure*}
\includegraphics[width=\textwidth]{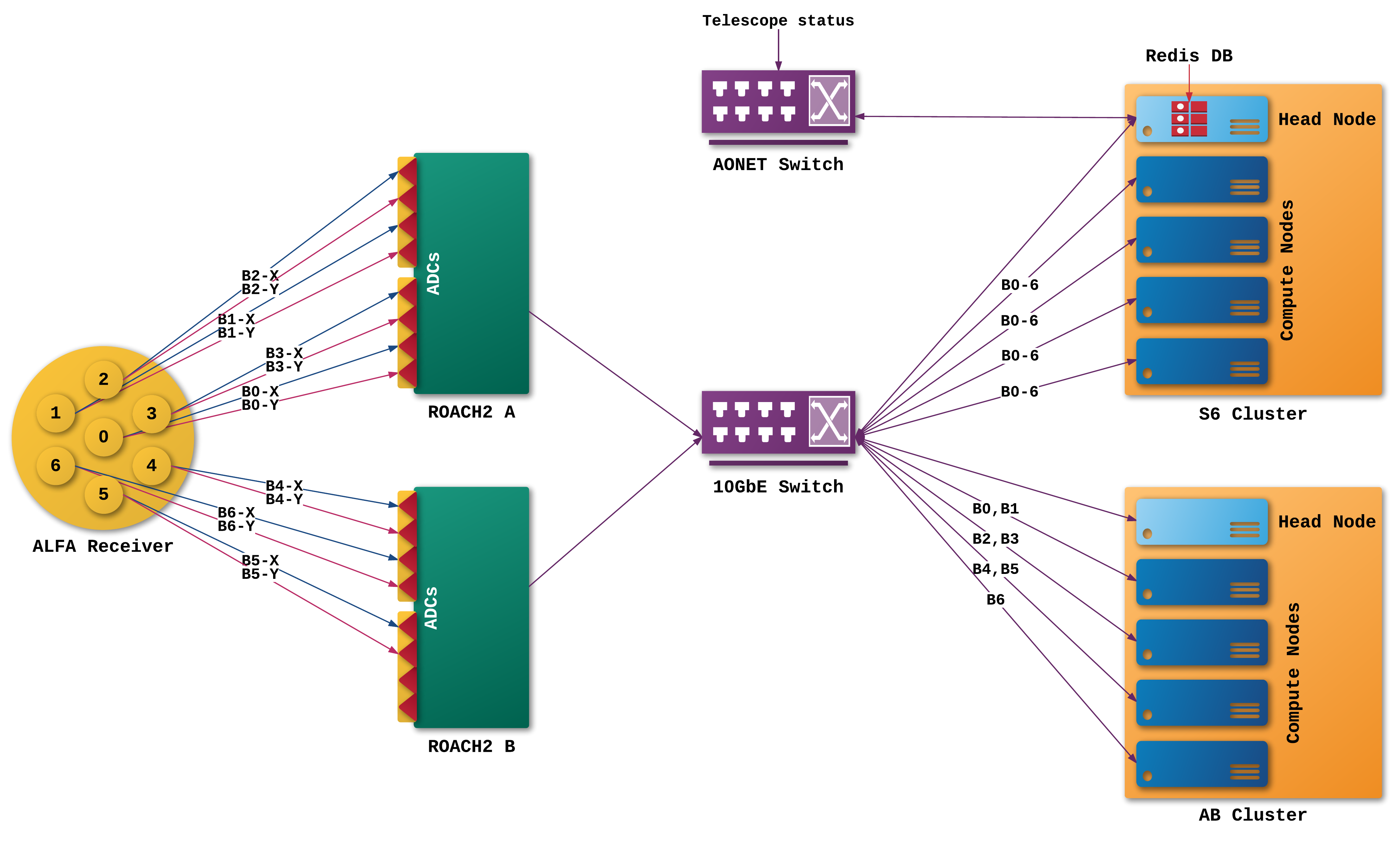}
\caption{Simplified high-level architecture of the SETIBURST system. Both
    polarizations (denoted by $X$ and $Y$) of all ALFA beams (denoted by
    $B0-B6$) are processed by two ROACH2 FPGA boards and distributed to compute
    nodes through a 10GbE swith. `AB' stands for ALFABURST in this diagram.
    The setup is described in detail in
    \S\ref{sec_sysdesc}.
    \label{fig_sysarch}}
\end{figure*}

Figure~\ref{fig_sysarch} shows the high-level architecture of the system.
The digital system processes signals from the ALFA receiver. ALFA is a
seven-beam system that operates in the 1225--1525~MHz range, with the seven
beams arranged in a hexagonal pattern. Each beam is approximately 3.5$'$ wide.
The receiver has a cold sky system temperature of $\sim$30~K. The central beam
has a gain of $\sim$11~K~Jy$^{-1}$, with the
peripheral beams having a slightly lower gain. Some of the sidelobes of the
ALFA beams are sensitive as well, with the peak of these sidelobes having a
loss of only $-$8.5~dB, i.e., the gain at the sidelobe peak is only (1/7)th of
that at boresight of the central beam. The advantage of this is that it has the
effect of increasing the area on the sky, and the disadvantages include
increased uncertainty in localization and poor fidelity in the measurement of
the spectral indices of FRBs that may have been picked up in these sidelobes
\citep{spi2014}. S6 uses 280~MHz of ALFA's 300~MHz bandwidth, while ALFABURST, in
its current version, supports a bandwidth of 56~MHz.

\subsection{FPGA Gateware}\label{sec_fpgadesc}

The 14 intermediate frequency (IF) signals from ALFA (one per polarization per
beam) are split before being distributed to various backends at the
observatory. SETIBURST hardware taps into these split signals, and digitizes
them. The hardware consists of two
ROACH2\footnote{`Reconfigurable Open Architecture Computing Hardware 2';
    \url{http://casper.berkeley.edu/wiki/ROACH2}} FPGA boards, each equipped
with two 1~Gsps
ADC16x250-8\footnote{\url{http://casper.berkeley.edu/wiki/ADC16x250-8}}
Analog-to-Digital Converters (ADCs). The first board (denoted by `ROACH2 A' in
Figure~\ref{fig_sysarch}) processes beams 0 through 3, while the second ROACH2
board (`ROACH2 B') processes beams 4 through 6. The ADCs first sample the data
at 896~MHz and digitizes it to 8 bits. The FPGA gateware uses a polyphase filter
bank (PFB) to channelize the data to 4096 channels, with a resulting time
resolution of 9.143~$\mu$s. The PFB is implemented using the standard
CASPER\footnote{Collaboration for Astronomy Signal Processing and Electronics
Research: \url{https://casper.berkeley.edu}} blocks
\texttt{pfb\_fir\_real}\footnote{\url{https://casper.berkeley.edu/wiki/Pfb_fir_real}}
and
\texttt{fft\_wideband\_real}\footnote{\url{https://casper.berkeley.edu/wiki/Fft_wideband_real}}.
The prefiltering uses 4 taps, and the co-efficients are the product of a sinc
function and a Hamming window.

\begin{figure*}
\begin{center}
\includegraphics[width=\textwidth]{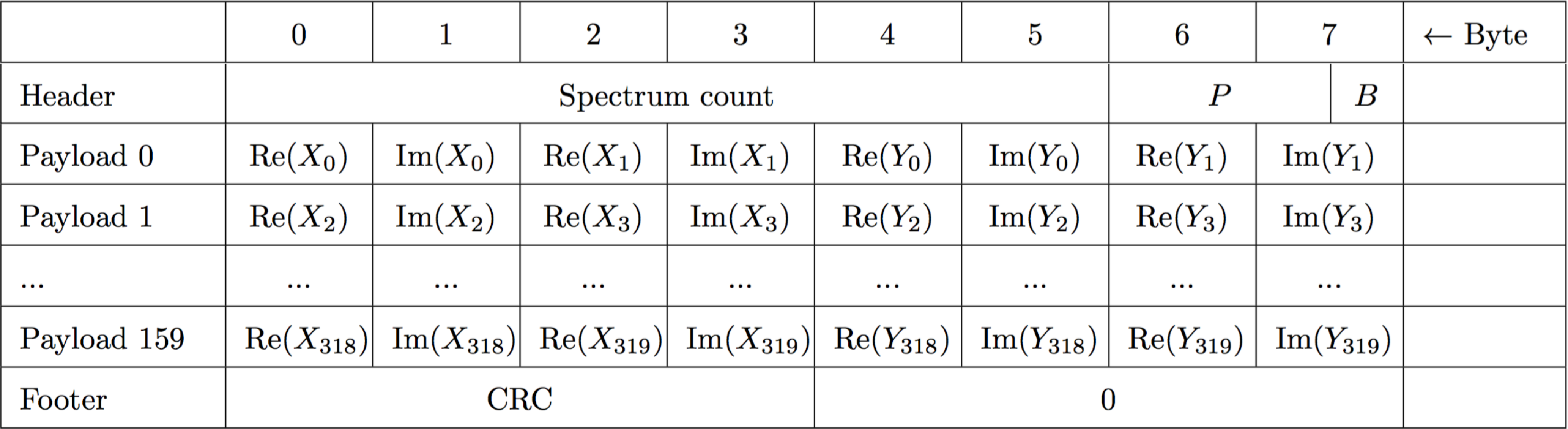}
\end{center}
\caption{The SERENDIP VI packet format, as described in \S\ref{sec_fpgadesc}. Each UDP
    payload is 1296 bytes long, with an 8-byte header and an 8-byte
    footer. The 12-bit field denoted by $P$ indicates the first channel of this
    packet, and the 4-bit field denoted by $B$ contains the beam identifier.
    Bytes are transmitted in left-to-right, top-to-bottom
    order, i.e., in network byte order.
\label{fig_s6pkt}}
\end{figure*}

The bandpass is split into eight sub-bands, which are packetized
separately, each addressed to a different HPC pipeline, and transmitted over
10-Gigabit Ethernet (10GbE). The packetization mechanism assigns a different IP
address to packets sent to different HPC nodes, using IP addresses stored in
software registers on the ROACH2 boards, that are programmable at run-time.
Each HPC node runs two software
instances/pipelines for all beams and polarizations (see \S\ref{sec_s6desc}).
Even though our digitization scheme results in a bandwidth of 448~MHz, ALFA has
a bandwidth of only 300~MHz. To remove channels with no information and to
reduce the output data rate, we pare the band down to 2560 channels
(corresponding to a bandwidth of 280~MHz) and these are packetized. The complex
samples at the output of the PFB are packetized into
1296-byte-long User Datagram Protocol (UDP) packets. Fig.~\ref{fig_s6pkt}
shows the S6 packet format. Each packet contains an 8-byte header that contains
a 48-bit spectrum counter, a 12-bit field indicating the first channel in the
packet (denoted by $P$ in Fig.~\ref{fig_s6pkt}), and a 4-bit beam identifier
that takes on values in the range 0--6 (denoted by $B$ in
Fig.~\ref{fig_s6pkt}). The spectrum count is used at the receiver
(HPC) for packet loss checking. Each packet also consists of a 64-bit footer
that contains a 32-bit cyclic redundancy check (CRC) for error detection on the
HPC. The bytes that make up these packets are transmitted in network
byte order.

\begin{figure*}
\begin{center}
\includegraphics[width=\textwidth]{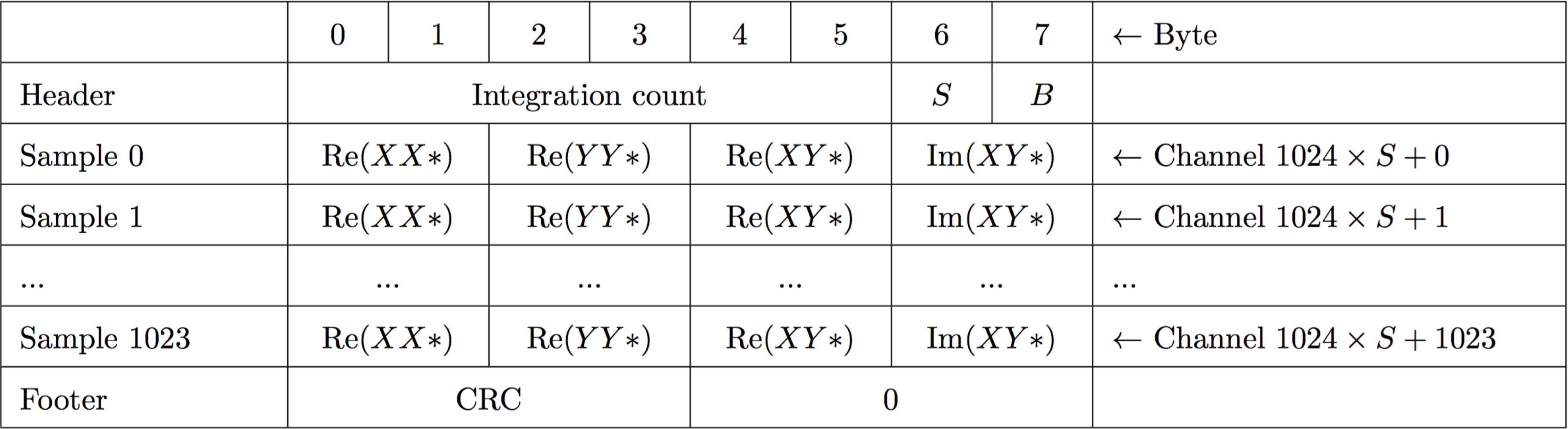}
\end{center}
\caption{The ALFABURST packet format, as described in \S\ref{sec_fpgadesc}.
    Each UDP payload is 8208 bytes long, with an 8-byte header and an 8-byte
    footer. The integration count and the sub-band identifier, denoted by
    $S$, together make it possible to check for missing packets. $B$ is the
    beam identifier. Bytes are transmitted in left-to-right, top-to-bottom
    order, i.e., in network byte order.
\label{fig_abpkt}}
\end{figure*}

For ALFABURST, polyphase channelization is followed by computation of
pseudo-Stokes values $XX^*$, $YY^*$, Re($XY^*$) and Im($XY^*$) (each 16 bits
wide), where $X$ and $Y$ are the Fourier representations of the two polarizations, and $X^*$ and
$Y^*$ are their respective complex conjugates. Note that full-Stokes values can
be computed from these values. The spectra are then time-integrated by a factor of
14, with a resulting time resolution of 128~$\mu$s. The spectra are then packetized
into UDP packets, and transmitted over 10GbE. To conform to the Ethernet
`jumbo frame' standard, each packet needs to be not more than 9000 bytes long.
Therefore, for each beam, the 4096-channel spectrum is split into four
sub-bands, each containing 1024 channels, with one sub-band per packet.
Fig.~\ref{fig_abpkt} shows the ALFABURST packet format. In
addition to the data, the UDP payload also contains a 64-bit header that
includes a 48-bit integration count, a one-byte sub-band identifier that takes
on values in the range 0--3 (denoted by $S$ in Fig.~\ref{fig_abpkt}), and a
one-byte beam identifier that takes on values in the range 0--6 (denoted by $B$
in Fig.~\ref{fig_abpkt}). The integration count, along with the sub-band
identifier allows us to check for missing packets on the receiving (HPC) side.
The integration count, along with a timestamp of when that count was
reset to zero -- which is read from elsewhere by the HPC software -- allows us
to get the timestamp of each packet. Each packet also consists of a footer
similar to the one in S6. As in the case of S6, bytes are transmitted in
network byte order.

\subsection{SERENDIP VI Software}\label{sec_s6desc}

The UDP packets that are transmitted by the FPGA are forwarded to appropriate
nodes in the HPC cluster by a Juniper Networks EX4500-LB 10GbE switch. The S6 HPC cluster
consists of five server-class computers -- one `head node', and four `compute
nodes', as shown in Figure~\ref{fig_sysarch}. Each compute node is equipped
with two Mellanox MCX312A-XCBT 10GbE network interface cards (NICs) that receive data from the
10GbE switch. Each compute node is a dual-socket, dual-GPU machine equipped
with RAID data disks. We use commercial gaming GPUs, namely NVIDIA GeForce GTX
780 Ti cards.

S6 uses the HASHPIPE\footnote{High Availability Shared Pipeline Engine;
    Available upon request.} software for data
acquisition and processing. HASHPIPE is
a multi-threaded data transport framework that moves high-bandwidth input data
from the 10GbE NICs through a series of shared memory ring buffers and
signal-processing threads, all the way to writing the output to
disk. HASHPIPE is designed with a modular architecture so that user-supplied
modules may be plugged in to perform certain tasks. The first module that
interfaces with the NIC is the `network thread'. The network thread reads data
from the NIC, checks for missing packets, and writes the data to the first
shared memory ring buffer. The next thread, the `GPU thread',
reads that data and performs fine channelization. Each of the 4096 channels
that arrive from the FPGA are channelized into 131072 channels by the GPU, with
a resulting frequency resolution of $\sim$0.8~Hz and time resolution of
1.198~s. This data is then written to the next shared memory ring buffer where
it is read by the next thread, the `CPU thread'. The CPU thread
performs thresholding as follows: To estimate the mean power level of spectra,
this thread boxcar averages each spectrum with a window of length 1024
channels, and computes the mean of the smoothed spectrum. Channels that have
values 20 times the mean are considered events of interest, which we term
`hits'. We note that the power spectra follow a $\chi^2$ distribution with 2
degrees of freedom, such that the mean and root mean square (RMS) are equal.
Therefore, the ratio of the detected power in a given channel to the mean power
level is a measure of $S/N$ in that channel. The hits are stored as a
function of time, frequency, and sky coordinates in a FITS file on disk.

\begin{figure*}
\begin{minipage}{0.5\textwidth}
\begin{center}
\includegraphics[width=\textwidth]{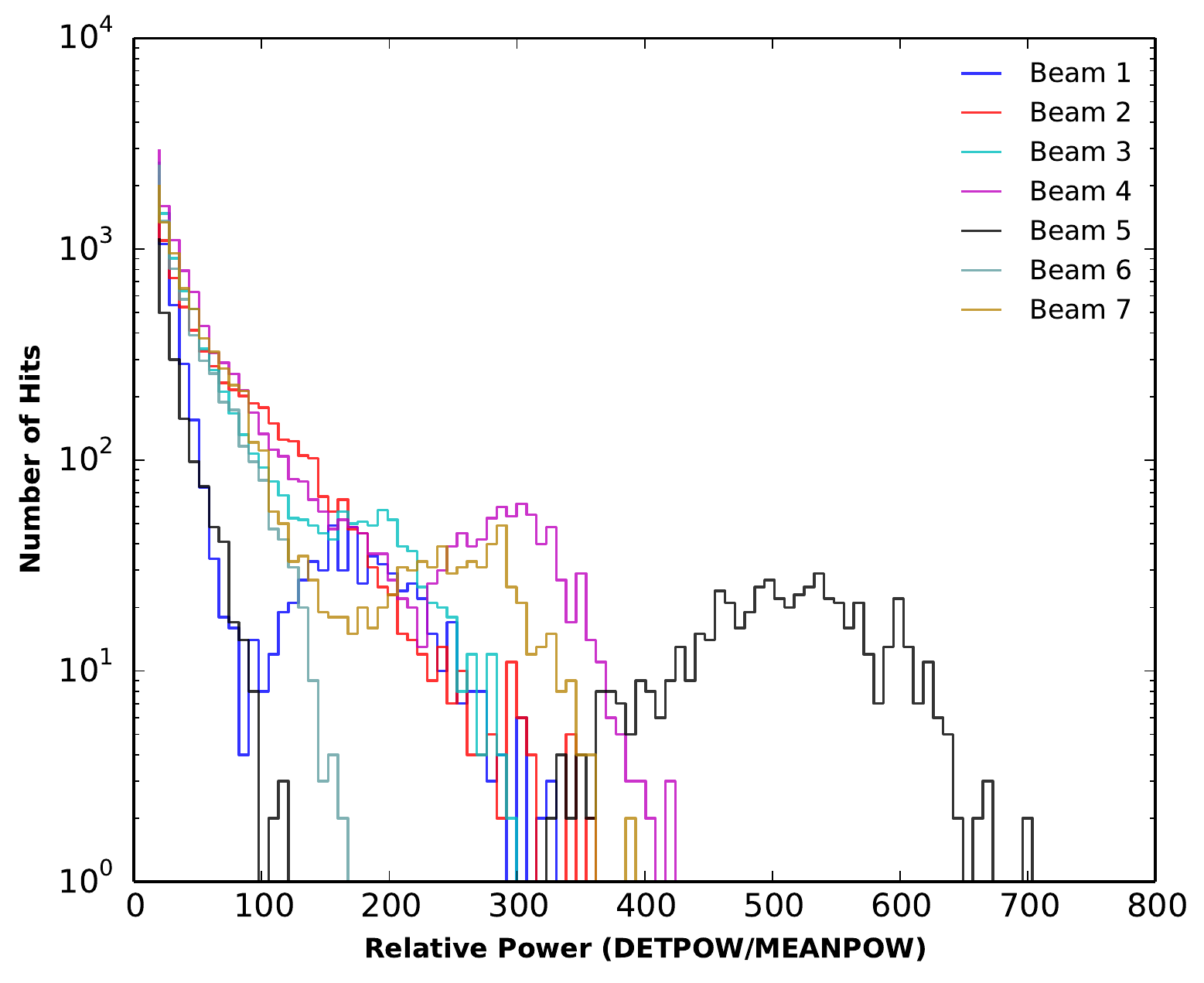}
\end{center}
\end{minipage}
\begin{minipage}{0.5\textwidth}
\begin{center}
\includegraphics[width=\textwidth]{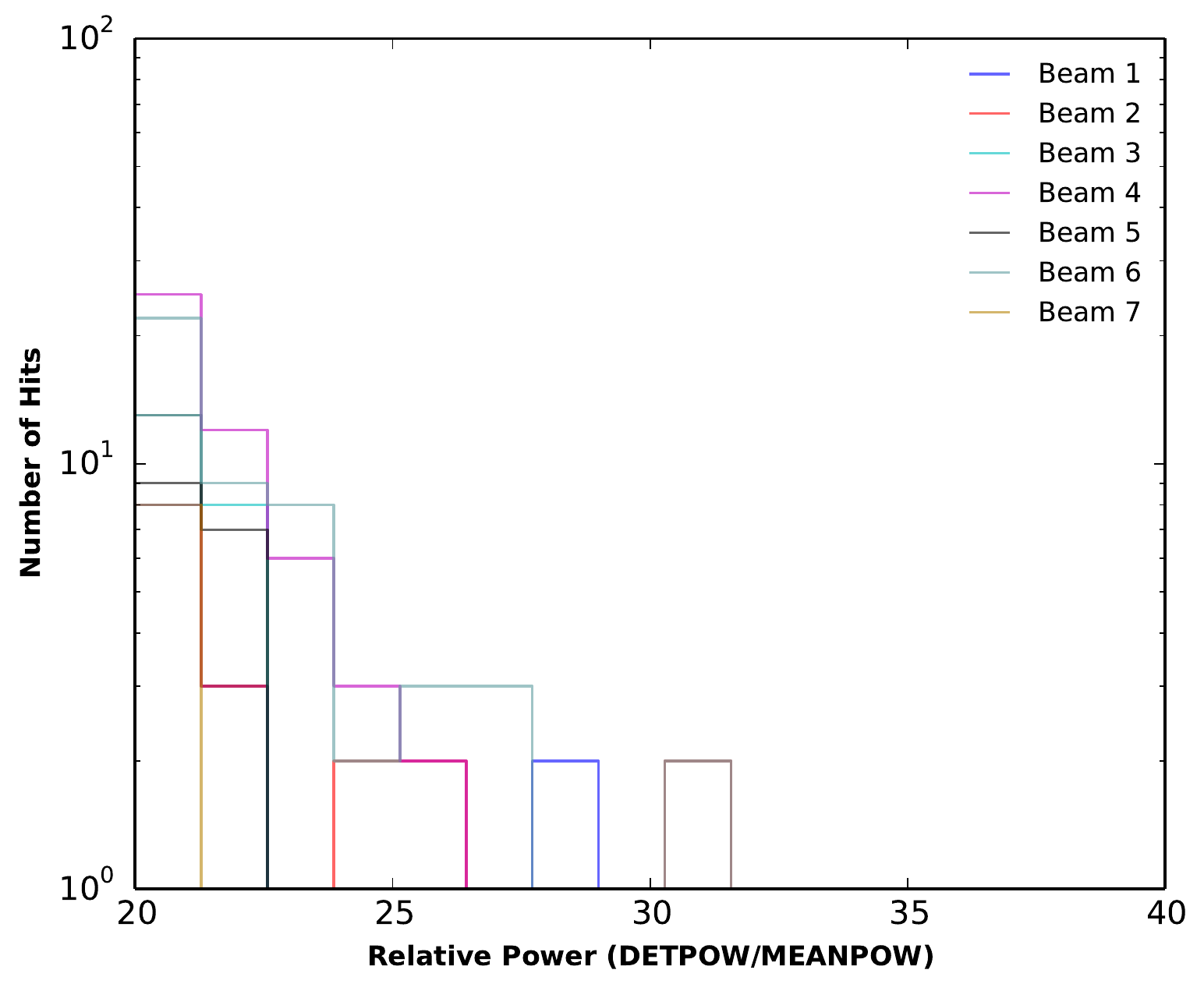}
\end{center}
\end{minipage}
\caption{Histograms of the ratio of detected power (`DETPOW') and mean
power (`MEANPOW') in a coincidence rejection comparison block (see \S\ref{sec_s6desc})
of a typical SERENDIP VI multi-beam data. This
    particular data comes from one of the system's 34~MHz sub-bands processed
    by a single pipeline instance, centred at 1252~MHz, spanning $\sim$800 seconds.
    The left panel shows the power distribution before RFI rejection, and the
    right panel shows the power distribution after the application of
    multi-beam RFI rejection that removes events that are found in two or more
    beams that are similar in frequency and time.\label{fig_s6plot}}
\end{figure*}

S6 also utilizes multi-beam coincidence RFI rejection. Individual hits from one
beam are checked for coincidence hits in the other beams. If hits are
found within 25000 frequency channels (corresponding to $\sim$20.9~kHz) on
either side of the event's channel, and five samples (corresponding to
5.99~s) on either side in time, across two or more beams, they are flagged as
RFI. These frequency and time spans, and our $S/N$ of 20, are chosen
empirically based on the prevailing average RFI conditions on site and our need
to ensure that potential astrophysical signals are retained as hits. Assuming
an exponential distribution of hit $S/N$ giving us
$n_\nu n_{\rm t} n_{\rm b} n_{\rm p} e^{-20}$ events, where $n_\nu$ is the number of
channels in our range of interest, $n_{\rm t}$ is the number of time samples,
$n_{\rm b}$ is the number of beams, and $n_{\rm p}$ is the number of
polarizations, for a $S/N$ of 20, we expect to detect $\sim$0.01 events in each
block used in the coincidence rejection comparison. In reality, however, the
statistics are dominated by RFI, and the actual number of detected events is
much larger. Figure~\ref{fig_s6plot} shows the performance of this RFI
rejection technique, applied to one polarization in one pipeline instance. We
note that these plots reflect a work in progress, and additional techniques
necessary to identify {\it{bona fide}} candidates are discussed elsewhere (Gajjar et~al., in preparation).

In addition to the signal processing software that runs on the compute nodes,
the S6 HPC system maintains a Redis key-value store on the head node. This is a
database that is constantly updated with the status of the telescope, read from
the Arecibo observatory's network, through a separate network switch (termed
`AONET switch' in Figure~\ref{fig_sysarch}). The information maintained in the
database includes the receiver in use, IF frequencies, and pointing and timing
information, among others.

As part of deployment and commissioning, we conducted various tests to verify
the functioning of the system. The primary end-to-end test involved injecting a
test signal with bandwidth $<$~0.8~Hz into the IF, and recovering that signal at
the expected level in the output data.

\subsection{ALFABURST Software}\label{sec_abdesc}

The ALFABURST HPC cluster is very similar to that of S6. It is made up of one head
node and four compute nodes. The main differences are that each compute node
is equipped with a single Mellanox MCX312A-XCBT 10GbE NIC that receives
UDP packets from the 10GbE switch, and uses NVIDIA GeForce GTX TITAN
GPU cards that have a larger memory than those used in S6, necessitated by
ALFABURST signal processing requirements.

The ALFABURST head node queries the S6 Redis database at a cadence of once per
minute, checking which receiver is at the focus. When an observer selects the
ALFA receiver, the corresponding value is updated in the Redis database. This
is detected by ALFABURST and data acquisition is initiated. While observation
is in progress, the head node continues to query the database for changes to
telescope state. Data acquisition is terminated when ALFA stops being the
selected receiver.

The compute nodes run the ALFABURST software data acquisition pipeline
instances.
Three of the compute nodes process data from two ALFA beams each, and
therefore, run two instances of the software. The remaining node processes the
seventh ALFA beam, and runs a single instance of the pipeline.
The software architecture follows 
a client-server model, where the server receives incoming data from the 10GbE
NIC, fills data corresponding to missing packets with zeros, and forwards the
data to the client. The client is modular by design, with each module handling
one logical signal processing stage. The ALFABURST data transport framework and
signal processing
system\footnote{The software is available upon request.} are based
on the ARTEMIS fast transient
search software developed for a recently-concluded survey at the UK station of
the LOFAR telescope \citep{kar2015}. Even though the software serves the
\cite{kar2015} survey sufficiently, the specifications of ALFABURST are much more
stringent, with a larger number of channels and a much larger bandwidth. The
software in its current form is not designed to process the entire ALFA
bandwidth at the native time resolution provided by the FPGA gateware. We
therefore process only a bandwidth of 56~MHz, and integrate the incoming
spectra with a resulting time resolution of 256~$\mu$s. We note that the
narrowest known FRB has a width of 350~$\mu$s \citep{pet2016}, so the choice of
time resolution is reasonable in this regard.

The first stage in the signal processing pipeline is the computation of
full-Stokes values from the pseudo-Stokes values in the packets following
\begin{eqnarray*}
    I &=& XX^* + YY^*,\\
    Q &=& XX^* - YY^*,\\
    U &=& 2{\rm Re}(XY^*),{\rm~and}\\
    V &=& -2{\rm Im}(XY^*).
\end{eqnarray*}
The search process requires only the total power (Stokes $I$), but it is worth
saving the other Stokes parameters for polarization studies of any detected
FRB. In our current implementation, we do not store the other Stokes
parameters, but this will be supported in future versions of the software.
Stokes computation is followed by the signal processing stages involved in
searching for FRBs. The following discussion briefly describes these signal
processing steps; for details, we refer the reader to \citet{kar2015}.

Since searching for FRBs involves correcting for the unknown dispersion delay
introduced by the ISM, the major signal processing operation involves
dedispersing the data over a range of trial DMs. We call this process the
dedispersion transform, converting dynamic spectra (frequency versus time
versus power) to a set of dedispersed time series (DM versus time versus
power). Dedispersion involves summing the data over all frequency channels, so
it is important to remove the data of strong RFI
that would otherwise result in a large number of false positives. Accordingly,
the data goes through the RFI clipper module. The RFI clipper implements an
adaptive thresholding algorithm \citep{kar2015} that takes into account the non-flat nature of
the bandpass and the time-varying baseline, and normalizes the output to have a
mean of 0 and a standard deviation of 1.

Following RFI removal and spectrum normalizaton, the data undergoes the
dedispersion transform. Being the most compute-intensive operation, this is
run on the GPU, and is implemented in Compute Unified Device Architecture
(CUDA)\footnote{\url{http://www.nvidia.com/object/cuda_home_new.html}}. The
dedispersion module is based on the Astro-Accelerate code developed by
\citet{arm2012}. The dedispersion transform is performed on data resident in
buffers 32768 samples long. At the currently-used time resolution of
256~$\mu$s, this corresponds to a duration of $\sim$8.4 s.
Temporal continuity across buffers is maintained by reusing
the last $n_{\rm maxshift}$ time samples from the previous buffer, where
$n_{\rm maxshift}$ is the number of time samples in the lowest frequency
channel to be `shifted', corresponding to the maximum DM.
If downstream
processing results in a detection, data from the buffer which contains the
pulse is saved to disk for later inspection.

We perform an incoherent dedispersion search over a DM range of
[0, 2560]~cm$^{-3}$~pc. The maximum DM among all known FRBs is
$\sim$1629~cm$^{-3}$~pc \citep{cha2016}, so our upper limit is a reasonable
choice. Even though the optimal DM step size is non-uniform across the DM range
of interest \citep{cor2003}, it is simpler to implement a fixed step size as is
done in \citet{arm2012}, thereby oversampling the DM space at larger values,
and, depending upon the step size, possibly undersampling the DM space at
smaller values. Undersampling the DM space implies the use of trial DMs that
are offset from the true values, and this leads to pulse broadening, resulting
in a loss in sensitivity. The smallest step size in an incoherent dedispersion
search required to minimize this loss in sensitivity is
\begin{equation}
\delta{\rm DM} = 1.205 \times 10^{-7}~t_{\rm samp} \nu^3 / \Delta\nu~{\rm
cm}^{-3}~{\rm pc},
\end{equation}
where $t_{\rm samp}$ is the sampling interval in ms, $\nu$ is the center
frequency in MHz, and $\Delta\nu$ is the bandwidth in MHz \citep{lor2005}. For
a nominal center frequency of 1375~MHz, our experimental setup yields
$\delta{\rm DM} \approx 1.4$~cm$^{-3}$~pc. As our fixed step size, we have
chosen 1~cm$^{-3}$~pc, which results in no loss in sensitivity.

\begin{figure}
\begin{center}
\includegraphics[height=0.5\textheight]{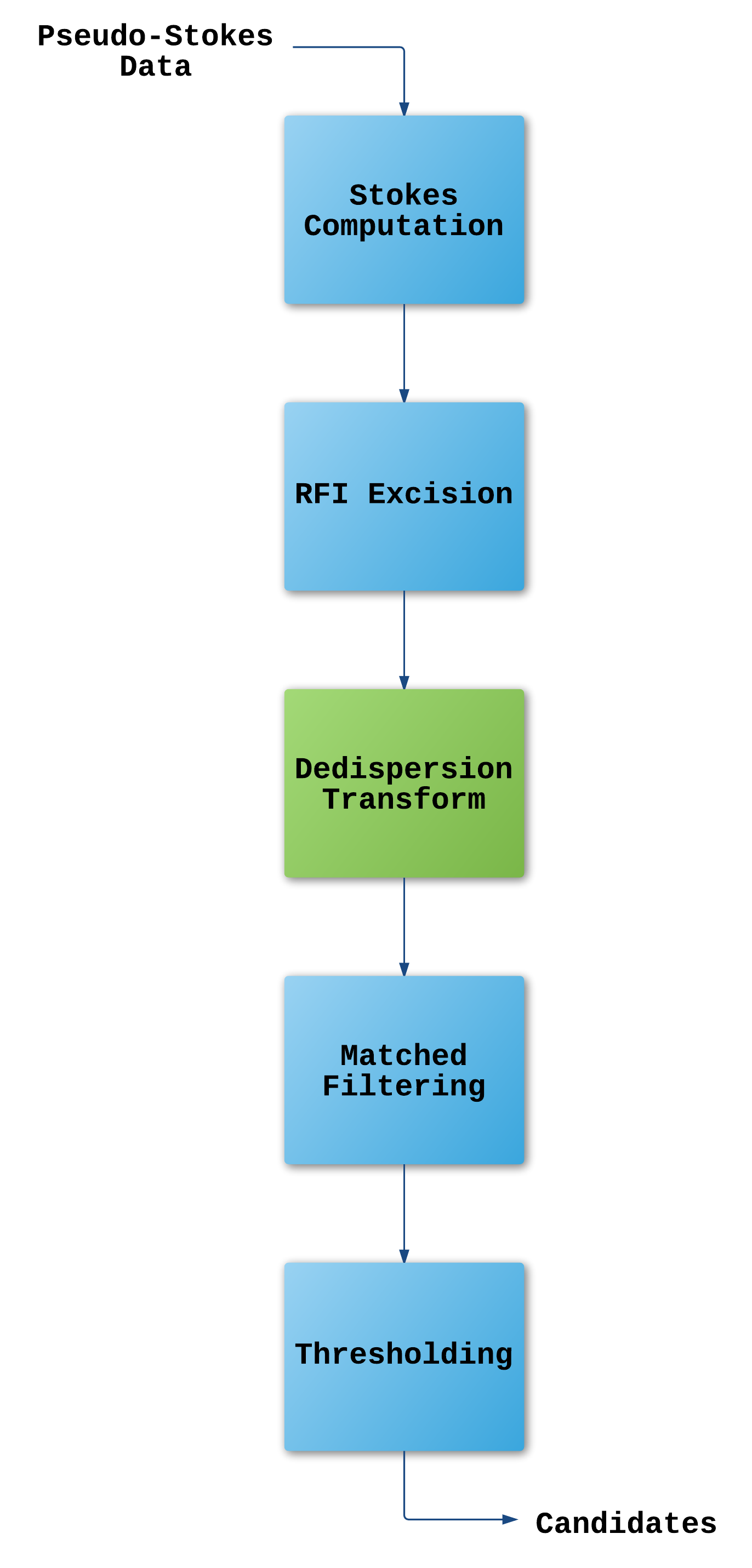}
\end{center}
\caption{Signal processing stages of the ALFABURST software pipeline. The dedispersion
    transform is implemented on GPUs.
    \label{fig_blockdiag}}
\end{figure}

The dedispersion transform step is followed by smoothing of the dedispersed
time series. Each time series is decimated by factors of 2 to 16, in powers of
2. This is a matched filtering operation meant to increase the sensitivity of
the search to pulses of varying widths. We note that the decimation is
performed block-wise in the current implementation, as opposed to using a
running window. Compared to doing true matched filtering, i.e., using a
running window, this has the effect of reducing the net sensitivity by a factor
of $\sqrt{2}$ \citep{kea2015}. All time series are then subject to a sensitivity threshold of
10 times the noise RMS. We do not
explicitly compute the RMS of the data. By design, the RFI clipper outputs spectra with
a standard deviation of 1, so we take the RMS to be the square root of the number of
channels that are summed during dedispersion. Events that cross the threshold are saved
to a candidates list that is written to disk, along with the RFI-removed
filterbank data corresponding to the data buffer the event was found in.
Fig.~\ref{fig_blockdiag} is a schematic of the aforementioned operations.

Observations are followed by the generation of diagnostic plots of
threshold-crossing event $S/N$ as a function of time and DM.
In the current scheme,
plots are automatically generated once a day, around noon local time. This makes plots
available within a few hours of recording of the signal. Web
pages containing these plots are automatically generated and made available
using a web server\footnote{\url{http://www.naic.edu/~alfafrb/}}. Beyond this
stage, data analysis is manual in nature. Plots are inspected visually, and
interesting events are followed up by examining the saved filterbank data. For
pulses that are seen in the filterbank data, we compare the pointing
information and DM to the entries in the ATNF Pulsar
Catalogue\footnote{\url{http://www.atnf.csiro.au/people/pulsar/psrcat/}}
\citep{man2005} to check whether they correspond to known pulsars.

\begin{figure*}
\begin{center}
\includegraphics[width=\textwidth]{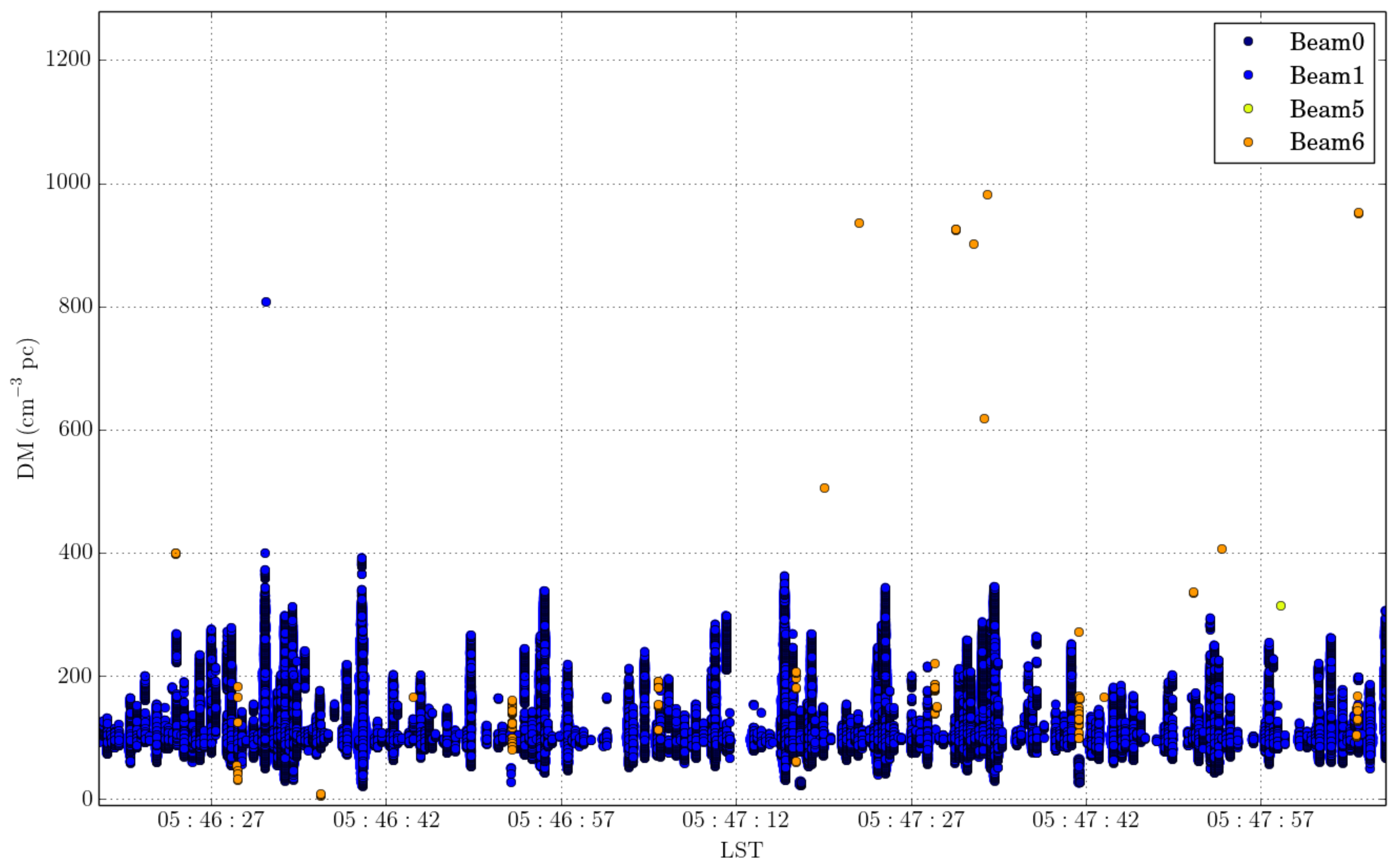}
\end{center}
\caption{ALFABURST commissioning test results for an observation of the pulsar
    B0611+22. The pointing was such that the pulsar was in beam 1 of ALFA.
    The markers represent events whose $S/N$ crossed our detection threshold.
    The clusters of detections are centered around a DM of 97~cm$^{-3}$~pc, as
    expected for this pulsar.
    \label{fig_7beam}}
\end{figure*}

Commissioning tests of the system were conducted from March to August 2015.
Fig.~\ref{fig_7beam} shows the results of one of our commissioning
observations, wherein we observed the pulsar B0611+22 in beam 1.
We obtained detections whose $S/N$ peaked around 97~cm$^{-3}$~pc, as expected
for the test pulsar. To verify the functionality of the system further, we
compared the number of events we detected to that obtained by applying the same
$S/N$ threshold to a time series that was dedispersed using the
SIGPROC\footnote{\url{http://sigproc.sourceforge.net/}} software, which is a
standard pulsar data processing package, yielding a match.

\section{Commensal Surveys}\label{sec_surv}

\subsection{Sky Coverage}\label{sec_skycov}

The S6 and ALFABURST surveys piggyback on ongoing Arecibo surveys, specifically the
PALFA and AGES surveys. PALFA is a survey for pulsars and fast transients
\citep{cor2006}, and has so far resulted in the discovery of 145 pulsars and
one FRB \citep{laz2015,spi2014}. Being a pulsar survey, PALFA
emphasizes coverage of the Galactic plane. PALFA pointings are towards the
inner Galaxy ($32^\circ \lesssim l \lesssim 77^\circ$; $|b| < 5^\circ$) and the
outer Galaxy ($168^\circ \lesssim l \lesssim 214^\circ$; $|b| < 5^\circ$),
with dwell times of 268~s and 180~s, respectively. PALFA has been allocated
230~hr. over the coming year.

AGES is an extragalactic HI survey, observing multiple fields spread in right
ascension across the northern sky \citep{aul2006}. Most fields are about
5$\times$4~sq.~deg. in size. Each pointing in the survey has a dwell time of
300~s. AGES is expected to observe for about 350~hr. over the coming year
(R. Minchin, private communication). AGES pointings are mostly away from the
Galactic plane, and therefore, are conducive to a search for extragalactic
FRBs.

\subsection{SERENDIP VI Sensitivity}\label{sec_s6surv}

For a narrow band signal in a single polarization, the minimum detectable flux,
in W~m$^{-2}$, of a signal that has width less than the channel bandwidth
$\Delta f$ Hz is given by
\begin{equation}
F = \sigma S_{\rm sys} \sqrt{\frac{\Delta f}{t}},
\end{equation}
where $\sigma$ is the threshold S/N, $S_{\rm sys}$ is the system equivalent
flux density (SEFD) in Jy, and $t$ is the integration time in s.

We use a detection threshold of 20-$\sigma$, which has been determined
empirically, based on the RFI environment at the site. Given that the boresight
SEFD of ALFA is 2.73 Jy, for our threshold S/N, for an integration time of
1.198~s, with a channel bandwidth of $\sim$0.8~Hz, the minimum detectable flux
is $\sim$4.6$\times10^{-25}$~W~m$^{-2}$, or $\sim$55~Jy across a
channel.

As an example of a transmitter, we consider the case of the 2380~MHz
transmitter of the Arecibo Planetary Radar, which is frequently used to
determine the orbits of near-Earth asteroids. It has an EIRP of
$\sim$2$\times10^{13}$~W. Our sensitivity is high enough to detect
similar transmitters up to a distance of $\sim$60~pc. However, this energetics
comparison is strictly for illustrative purposes. The detectability and
decoding of terrestrial analogs at interstellar distances is a complex topic
\citep[see][]{sul1978} and is not addressed here.

\subsection{ALFABURST Sensitivity}\label{sec_absurv}

Following the radiometer equation \citep[see, for example,][]{lor2005}, the
threshold flux density of a single pulse search,
\begin{equation}
S_{\rm min} = \frac{\sigma S_{\rm sys}}{\sqrt{n_{\rm p} \Delta
        f W}},
\end{equation}
where $\sigma$ is
the threshold S/N, $S_{\rm sys}$ is the SEFD, $n_{\rm p}$ is the number of
polarizations, $\Delta f$ is the bandwidth in MHz, and $W$ is the pulse width
in ms.
In the absence of RFI, the choice of
$S/N$ threshold
for a single pulse search is
rather straightforward. If we assume Gaussian statistics, the number of events
crossing the threshold $\sigma$ is
\begin{equation}
N(>\sigma) \approx 2 n_{\rm samp} \theta n_{\rm DM},
\end{equation}
where $n_{\rm samp}$ is the number of time samples, $n_{\rm DM}$ is the number
of DM channels, and $\theta$ is the probability of occurrence of a sample with
peak above $\sigma$ \citep{cor2003}. A reasonable value of $\sigma$ for one ALFABURST
buffer, i.e., about 8.4 s, such that the number of events due to noise alone is
1, comes out to be 5.7. In practice, however, RFI poses a significant problem,
especially at Arecibo -- both due to the noisy environment and the high
sensitivity of the telescope -- and requires us to choose a threshold that is
much higher. We have empirically determined 10 as our optimal $S/N$ threshold,
eliminating a large fraction of spurious events, while minimizing the
likelihood of missing a potential astrophysical signal.
After applying this threshold, we discard less than 5 per cent of the observed
time span to RFI.

For an FRB with a width of 1~ms, located at the central beam boresight of ALFA,
our sensitivity is $\sim$2.6~mJy. All known FRBs have observed peak flux
densities ranging upwards of $\sim$200~mJy, therefore, in spite of not
utilizing the whole ALFA band, our sensitivity is reasonable.

\subsubsection{Event Rates}\label{sec_abdisc}

Given that the ALFABURST survey is a commensal survey that piggybacks on
multiple surveys intended for multiple applications, with each survey observing
a different part of the sky for different amounts of time, and the fact that
FRB event rates and Galactic latitude dependence are not well constrained, it
is challenging to come up with a rigorous expectation of the number of
detections. Therefore, we follow a naive approach, merely extrapolating from
the expected usage duration and instantaneous field-of-view (FoV) of ALFA, and
the event rate computed by \citet{sch2016} based on the Arecibo FRB detection.
ALFA is expected to be used by PALFA and AGES for $\sim$580~hr. over the next
year.
It has an instantaneous FoV of $\sim$0.02~sq.~deg. within the full-width
half-maximum. \citet{sch2016} compute an event rate of
$5.08^{+17.78}_{-4.81}\times10^4$~sky$^{-1}$~day$^{-1}$ for bursts with flux
density above 57~mJy. Together, this leads to an expectation of between 0 and 5
such FRB detections in the coming year.

\section{Conclusion}\label{sec_conc}

We have designed, built, and deployed an automated, commensal, realtime
multi-science backend for the Arecibo telescope that conducts two surveys
simultaneously. S6 is conducting a survey for technologically advanced
life, whereas ALFABURST is conducting a survey for fast radio transients.

Future work for S6 involves replicating the system at other observatories. We
are in the process of building a similar system for the Green Bank
Telescope. Part of the data from both Arecibo and Green Bank will eventually be
sent out over the SETI@home\footnote{\url{http://setiathome.berkeley.edu/}}
citizen science distributed computing system for processing.

Future work for ALFABURST involves supporting the whole ALFA bandwidth of 300~MHz, the
native FPGA time resolution of 128~$\mu$s, and searching a larger range of DMs.
The fact that ALFA has multiple beams can be used for the coincidence rejection
of RFI -- if a signal appears in all seven beams, it is likely that it is of
terrestrial origin. We also intend to support the 327~MHz receiver whose usage
is more than that of ALFA, letting us not only increase the survey time, but
also perform a realtime, commensal survey for FRBs at a relatively less
explored part of the spectrum. The results of a 327~MHz survey would enable us
to constrain the spectral index of FRBs, similar to what was done with the
non-detection at 145~MHz by \citet{kar2015}. Longer-term goals include
reducing the latency involved in the generation of diagnostic plots, automatic
classification of candidate signals, and implementing a mechanism for
triggering telescopes that operate at lower frequencies, following a detection.

\acknowledgments

We thank the anonymous referee for feedback that greatly improved the
manuscript.
We thank the staff of the Arecibo Observatory, in particular, Hector Hernandez,
Robert Minchin, Mike Nolan, Phil Perillat, Luis Quintero, Joan Schmelz, Arun
Venkataraman, and Dana Whitlow for support with instrument deployment and
commissioning tests. We also acknowledge the cooperation of the PALFA and AGES
consortia. JC, AK, and WA thank the Leverhulme Trust for support with
instrument development. DRL and MAM were supported through NSF Award \#1458952.
Some of the ALFABURST hardware was purchased with funds from the WVU Eberly College of
Arts and Sciences and the WVU Department of Physics and Astronomy. We also
thank the NASA ASTID Program, the John Templeton Foundation, and the
Breakthrough Prize Foundation for their support. We thank Eric Korpela for
useful discussions.

\facility{Arecibo}

\bibliography{ref}

\clearpage

\end{document}